\documentclass[%
 preprint,
 amsmath,amssymb,
 aps,
]{revtex4-1}
\usepackage{graphicx}
\usepackage{dcolumn}
\usepackage{bm}
\begin{document}


\title{Centrifugal force reversal from the perspective of rigidly rotating observer}


\author{Giorgi Dalakishvili}
\affiliation{Center for Theoretical Astrophysics, Institute of
Theoretical Physics,
  Ilia State University, Tbilisi, Georgia}



\begin{abstract}
In previous studies  the dynamics of the relativistic particle
moving along the rotating pipe was investigated. The simple gedanken
experiment was considered. It was shown that at large enough
velocities a centrifugal force acting on the bead changes its usual
sign and attracts towards the rotation axis. The authors studied
motion of the particle along the rotating straight pipe in the frame
of the observer located in the center of rotation, also dynamics of
centrifugally accelerated relativistic particle was studied in the
laboratory frame. In the both cases it was shown that centrifugal
force changes sign. Recently the problem was studied in the frame of
stationary observers. It was argued that centrifugal acceleration
reversal  is not frame invariant effect. In the present paper we
consider motion of particle along the rotating straight line in the
frame of an arbitrary stationary observer located at certain
distance form the center of rotation and rigidly rotating with
constant angular velocity. It is shown that any stationary observer
could detect reversal of centrifugal acceleration.
\end{abstract}


\maketitle



\maketitle

\section{Introduction}
In [1] the authors studied the two-dimensional, relativistic motion
of a particle along a rotating field line. In this study, the case
of a straight trajectory was considered. The model was highly
idealized and purely mechanical but it had the advantage of having
exact analytic solutions. Authors found that the particle does not
cross the ``light cylinder" radius $R_c$ (the light cylinder radius
is defined as a hypothetical surface where the linear velocity of
rotation equals the speed of light, $R_{c}\equiv{c}/\omega$, where
$\omega$ is the angular velocity of rotation and $c$ denotes the
speed of light). In this study it was also shown that if the initial
radial velocity of the particle $\upsilon_{r0}\geq{c}/\sqrt{2}$,
then the radial acceleration is initially negative (cf.\ the
special-relativistic effect of the ``centrifugal force reversal",
[1]).  In other words, the motion of a particle moving along a
rotating, straight linear ``pipe"  is limited by $R_{c}$, and during
the course of its motion the particle may experience not only
acceleration, but also deceleration. Behavior of centrifugal force in the
frame work of the general relativity was studied in [2] and [3], it was found that centrifugal force could change sign.
The main idea of the `pipe-bead' (`rotator-pipe-bead') \textit{gedanken} (see [4], [1],
[5], [6], [7]) experiments was to mimic the common situation in
relativistic and rotating astrophysical flows, where the plasma
particles are forced to move along the field lines of governing
magnetic fields.

Recently in [8] was argued that  the effect discussed by [1] is not
frame-invariant and disappears if one uses frame-invariant
quantities. Thus, in special relativity, there is no reversal of
centrifugal acceleration. The effect seen in [1] is a time dilation
([9]) and it describes an unphysical coordinate acceleration.

In [8] author raised interesting idea to study the centrifugal force
reversal in the frame of different observers. In the present study
the motion of the particle is considered in the of the observer
located at certain distance from the center of rotation and rigidly
rotating with constant angular velocity.

\section{Main consideration}

As in [1]) and [8] the problem is considered in geometrical units in
which $G=c=1$, while the metric signature is: $(-1,1,1)$.

Let us first consider the the dynamics of the particle moving along
the rotating wire in the frame of reference rotating with the
pipe-bead system (i.e.\ rotating frame - RF). In order to do this,
we first need to switch to the frame, rotating with the angular
velocity $\omega$. Employing the transformation of variables:
\begin{align}
\label{eq:1}
 T=t,\\
\label{eq:2}
 \tilde\varphi=\varphi+\omega t
\end{align}

one arrives to the metric:
\begin{equation}
\label{eq:3}
ds^{2}=-(1-\omega^{2}r^{2})dt^{2}+2\omega^{2}dtd\varphi+r^{2}d\varphi^{2}+dr^{2}.
\end{equation}

For the straight pipe $(\varphi=\varphi_{0}=const)$ case,
Eq.~\ref{eq:3} reduces to the metric $ds^{2} =
-(1-\omega^{2}r^{2})dt^{2}+dr^{2},$ which is the basic metric for
the [1]) study.

The resulting metric tensor
\begin{equation}
\label{eq:4a} \|g_{\alpha\beta}\|=
\left(%
\begin{array}{cc}
  -(1-\omega^{2}r^{2}) & 0 \\
  0 & 1 \\
\end{array}%
\right),\tag{4a}
\end{equation}
we find:
\begin{equation}
\label{eq:4b}
\Delta\equiv[-det(g_{\alpha\beta})]^{1/2}=\left(1-\omega^{2}r^{2}\right)^{1/2},\tag{4b}
\end{equation}

For this relatively simple, but non-diagonal, two-dimensional
space-time one can develop the ``$1+1$" formalism. In doing so we
can follow, as a blueprint, the well-known ``$3+1$" formalism,
widely used in the physics of black holes ([10], [11], [12]). Namely,
we first introduce the definitions of the \textit{lapse function}:
\begin{equation}
\label{eq:5}
\alpha\equiv\frac{\Delta}{g_{rr}}=\sqrt{1-\omega^{2}r^{2}},\tag{5}
\end{equation}

Within this formalism Eq.~\ref{eq:3} can be presented in the
following way:
\begin{equation}
\label{eq:6}
 ds^{2}=-\alpha^{2}dt^{2}+g_{rr}dr^{2}.\tag{6}
 \end{equation}

Note that for the metric tensor, Eq.~\ref{eq:4a}, $t$ is the cyclic
coordinate and, moreover, in the RF the motion of the bead inside
the pipe is geodesic, i.e.\ there are no external forces acting on
it. Hence the proper energy of the bead, $E_{0}$, must be a
conserved quantity. Employing the definition of the four velocity
$U^{\alpha}\equiv dx^{\alpha}/d\tau$ one can write:
\begin{equation}
\label{eq:7}
 E_{0}\equiv-U_{t}=-U^{t}g_{tt}=const.\tag{7}
\end{equation}
On the other hand, the basic four-velocity normalization condition
$g_{\alpha\beta}U^{\alpha}U^{\beta}=-1$ requires:
\begin{equation}
\label{eq:8a}
U^{t}=[-g_{tt}-g_{rr}\upsilon_{r}^{2}]^{-1/2}.\tag{8a}
\end{equation}
This equation, written explicitly, has the following form:

\begin{equation}
\label{eq:8b}
 U^{t}=[1-\omega^{2}r^{2}-\upsilon_{r}^{2}]^{-1/2}.\tag{8b}
\end{equation}

Recalling  the definition of the Lorentz factor $\gamma(t)={1 \over
\sqrt{1-r^2\omega^2-\upsilon_{r}^2}}$ , with $\upsilon_{r}=dr/dt$, one can
easily see that:
\begin{equation}
\label{eq:8c}
U^{t}=[1-r^{2}\omega^{2}-\upsilon_{r}^{2}]^{-1/2}=\gamma(t).\tag{8c}
\end{equation}
It is important to note that the conserved proper energy of the
bead, $E_{0}$, defined by Eq.~\ref{eq:7}, may be written simply as:

\begin{equation}
\label{eq:9} E_{0}=\gamma(t)[1-r(t)^{2}\omega^{2}]=const.\tag{9}
\end{equation}

From Eq.~\ref{eq:9} follows that radial velocity of particle is:

\begin{equation}
\label{eq:10}
\left(\partial_{t}r\right)^{2}=(1-\omega^{2}r^{2})\left(1-\frac{1-\omega^{2}r^{2}}{E_{0}^{2}}\right).\tag{10}
\end{equation}

Now let us consider motion of the test particle in the frame of the
observer, located at distance $R_{SO}$  from the center
of rotation and moving with velocity $\omega R_{SO}$ perpendicular to
$\overrightarrow{R_{SO}}$. Let us consider the case when azimuthal
coordinate if the observer coincides with azimuthal coordinate of
the test particle. The above introduced frame instantaneously is
equivalent to the inertial frame moving with velocity $\omega
R_{SO}$ perpendicular to $\overrightarrow{R_{SO}}$ at the point of
observer location. Eq.~\ref{eq:11} gives relation between time
interval ($dt$) measured by observer located at the center of
rotation and rotating with angular velocity $\omega$ and time
interval ($dt_{SO}$) measured by stationary observer located at
$R_{SO}$ and rotating with angular velocity $\omega$.
Eq.~\ref{eq:12} relates change of the test particle radial
coordinate ($dr$) measured by observer located at the center of
rotation and rotating with angular velocity $\omega$ and change of
particle radial coordinate ($dt_{SO}$) measured by stationary
observer located at $R_{SO}$ and rotating with angular velocity
$\omega$.

\begin{align}
  \label{eq:11}
 dt_{SO}=\sqrt{1-\omega^{2}R_{SO}^2}dt,\tag{11}\\
  \label{eq:12}
  dr_{SO}=dr.\tag{12}
\end{align}

We should take into account that radial coordinate of stationary
observer $R_{SO}$ is constant and does not change in time, while
radial coordinate of test particle is time dependent variable i.e.,

\begin{align}
  \label{eq:13}
 \frac{dR_{SO}}{dt}=\frac{dR_{SO}}{dt_{SO}}=0,\tag{13}\\
  \label{eq:14}
  \frac{dr_{SO}}{dt_{SO}}\neq 0,\tag{14}\\
\label{eq:15}
  \frac{dr}{dt}\neq 0.\tag{15}
\end{align}

Velocity of change of test particle radial coordinate o measured by
stationary observer could be expressed es follows:

\begin{equation}
\label{eq:16}
\left(\partial_{t_{SO}}r_{SO}\right)^{2}=\frac{1-\omega^{2}r_{SO}^{2}}{1-\omega^{2}R_{SO}^{2}}\left(1-\frac{1-\omega^{2}r_{SO}^2}{E_{0}^{2}}\right),\tag{16}
\end{equation}

when observer is located in the center of the rotation i.e
$R_{SO}=0$ then Eq.~\ref{eq:16} transforms to the following
expression:

\begin{equation}
\label{eq:17}
\left(\partial_{t}r\right)^{2}=(1-\omega^{2}r^{2})\left(1-\frac{1-\omega^{2}r^{2}}{E_{0}^{2}}\right),\tag{17}
\end{equation}

Eq.~\ref{eq:17} coincides with expression derived in [1]

When particle passes stationary observer located at $R_{SO}$ i.e
$R_{SO}=r_{SO}$, the Eq.~\ref{eq:16} transforms in the following
expression

\begin{equation}
\label{eq:18}
\left(\partial_{t_{SO}}r_{SO}\right)^{2}=\left(1-\frac{1-\omega^{2}R_{SO}^2}{E_{0}^{2}}\right),\tag{18}
\end{equation}

Eq.~\ref{eq:18} coincides with expression derived in [8].

From Eq.~\ref{eq:16} we could derive expression for test particle
radial acceleration measured by stationary observer:

\begin{equation}
\label{eq:19} \frac{\partial^{2} r_{SO}}{\partial
t_{SO}^{2}}=\frac{\omega^{2}r_{SO}^{2}}{1-\omega^{2}R_{SO}^{2}}\left(1-\frac{2\left(\partial_{t_{SO}}r_{SO}\right)^{2}\left(1-\omega^{2}R_{SO}^{2}\right)}{1-\omega^{2}r_{SO}^{2}}\right).\tag{19}
\end{equation}

If stationary observer is located in the center of rotation i.e.,
$R_{SO}=0$ than Eq.~\ref{eq:19} transforms to the following
expression:

\begin{equation}
\label{eq:20} \frac{\partial^{2} r}{\partial
t^{2}}=\omega^{2}r^{2}\left(1-\frac{2\left(\partial_{t}r\right)^{2}}{1-\omega^{2}r^{2}}\right),\tag{20}
\end{equation}

Eq.~\ref{eq:20} coincides with expression of acceleration derived in
[1].

When particle passes stationary observer located at $R_{SO}$ i.e,
$R_{SO}=r_{SO}$, the particle radial acceleration measured by
stationary observer is expressed as follows:

\begin{equation}
\label{eq:21} \frac{\partial^{2} r_{SO}}{\partial
t_{SO}^{2}}=\frac{\omega^{2}R_{SO}^{2}}{1-\omega^{2}R_{SO}^{2}}\left(1-2(\partial_{t_{SO}}r_{SO})^{2}\right).\tag{21}
\end{equation}

From Eq.~\ref{eq:21} we see that if particle passes observer with
radial velocity larger then $1/\sqrt{2}$ i.e,
$\partial_{t_{SO}}r_{SO}>1/\sqrt{2}$, then radial acceleration of
the test particle measured by stationary observer is negative, i.e.,
$\partial^{2} r_{SO}/\partial t_{SO}^{2}<0$. From Eq.~\ref{eq:19}
follows that when particle is located in the center of rotation its
radial acceleration measured by stationary observer is negative in
the case when
$2(\partial_{t_{SO}}r_{SO})^{2}(1-\omega^{2}R_{SO}^{2})<1$ i.e.,
velocity measured by stationary observer satisfies condition:

\begin{equation}
\label{eq:22}
(\partial_{t_{SO}}r_{SO})^{2}(1-\omega^{2}R_{SO}^{2})>1/2.\tag{22}
\end{equation}

After taking into account transformations Eq.~\ref{eq:11} and
Eq.~\ref{eq:12} we can see that condition given by Eq.~\ref{eq:22}
is satisfied the particle velocity measured by observer located in
the center of rotation satisfies following condition:
\begin{equation}
\label{eq:23}
\partial_{t}r>\frac{1}{\sqrt{2}}.\tag{23}
\end{equation}

\section{Conclusions}

In this study the dynamics of centrifugally accelerated relativistic
particle is studied in the frame rigidly rotating observers located
at certain distance from the center od rotation. The expressions for
the particle radial velocity and radial acceleration are derived. It
was analyzed whether any observer could detect reversal of the
centrifugal acceleration.

From Eq.~\ref{eq:19}, Eq.~\ref{eq:22} and Eq.~\ref{eq:23} follows
that radial acceleration of the test particle located in the center
of the rotation measured by any stationary observer is negative when
particle  radial velocity measured by observer located in the center
of rotation is larger than $1/\sqrt{2}$. From Eq.~\ref{eq:21}
follows that particle radial acceleration measured by stationary
observer \textit{at the moment when particle passes observer} is
negative when at this moment particle radial velocity measured by
stationary observer is larger than $1/\sqrt{2}$. These results are
in agreement with result obtained in [1]. We could conclude that
centrifugal acceleration reversal described in [1] occurs in the
frame of any stationary observer. \textit{When one needs to take
derivative of expression in respect of time, one should take into
account that radial coordinate of moving particle and observer
coincide at the particular moment of time and they are not equal for
arbitrary moment of time, in other word one should keep in mind that
radial coordinate of stationary observer does not depend on time
while the radial coordinate of particle varies in time. }


\nocite{*}
\bibliographystyle{apsrev4-1}
\bibliography{centrifugal}

\end{document}